\documentclass[epj]{svjour}
\usepackage{graphicx}
\usepackage{amssymb}

\newcommand*{\ep}{\epsilon}
\newcommand*{\de}{\delta}
\newcommand*{\om}{\omega}
\newcommand*{\si}{\sigma}

\newcommand*{\ua}{\uparrow}
\newcommand*{\da}{\downarrow}
\newcommand*{\eff}{\mathrm{eff}}
\newcommand*{\crit}{\mathrm{c}}
\newcommand*{\sgn}{\mathrm{sgn}}

\newcommand*{\abs}[1]{\left|#1\right|}
\newcommand*{\aver}[1]{\left<#1\right>}
\newcommand*{\averaa}[1]{\langle #1\rangle}
\newcommand*{\bra}[1]{\left<#1\right|}
\newcommand*{\ket}[1]{\left|#1\right>}
\newcommand*{\braaa}[1]{\langle#1|}
\newcommand*{\ketaa}[1]{|#1\rangle}

\begin{document}
\title{Fundamental properties, localization threshold, and the 
  Tomonaga--Luttinger behavior of electrons in nanochains}

\author{Adam Rycerz \and Jozef Spa{\l}ek}
\institute{Marian Smoluchowski Institute of Physics, \\
    Jagiellonian University, Reymonta 4, 30-059~Krak\'{o}w, Poland, \\
    \email{adamr@th.if.uj.edu.pl, ufspalek@if.uj.edu.pl}}
\abstract{
We provide a fairly complete discussion of the electronic properties of 
nanochains by modelling the simplest quantum nanowires within a recently 
proposed approach which combines the 
\emph{\textbf{E}xact \textbf{D}iagonalization} in the Fock space with 
\textit{\textbf{Ab} \textbf{I}nitio} calculations (EDABI method). 
In particular, the microscopic parameters of the second--quantized Hamiltonian
are determined, and the evolution of the system properties is traced in a 
systematic manner as a function of the interatomic distance (the lattice 
parameter, $R$). 
Both the many--particle ground state and the dynamical correlation functions 
are discussed within a single scheme.
The principal physical results show:
$(i)$ the evolution of the electron momentum distribution and its analysis 
in terms of the Tomonaga--Luttinger scaling,
$(ii)$ the appearance of mixed metallic and insulating features (\emph{partial
localization}) for the \emph{half--filled} band case, 
$(iii)$ the appearence of a universal \emph{renormalized} dispersion relation 
for the electron energy, which incorporates both the band--structure and the 
Hubbard--splitting features in the presence of electron interactions, and 
$(iv)$ the transformation from a \emph{highly--conducting} nanometallic 
state to the \emph{charge--ordered} nanoinsulator in the quarter--filled case.
The analysis is performed using the Wannier functions composed of an adjustable
Gaussian 1$s$--like basis set, as well as includes a \emph{long--range} 
part of the Coulomb interaction.
}
\authorrunning{A.\ Rycerz and J.\ Spa{\l}ek}
\titlerunning{Fundamental properties, localization threshold, and 
 	Tomonaga--Luttinger behavior ...}
\maketitle

\section{Introduction}
Recent developments in computational as well as analytical methods have lead 
to a~successful determination of the electronic properties of semiconductors 
and metals starting from LDA \cite{hoko}, LDA+U \cite{aniza}, and related 
\cite{svangu} approaches. 
Even strongly correlated systems, such as V$_2$O$_3$ (which undergoes the Mott
transition) and high--temperature superconductors have been treated in this 
manner \cite{ezho}.
However, the discussion of the metal--insulator transition of the 
Mott--Hubbard type is not yet possible in a systematic manner, particularly 
for low--dimensional systems. 
These difficulties are caused by the circumstance where the 
elec\-tron--electron interaction is comparable, if not stronger, than the 
single--particle energy. 
In effect, the procedure starting from the single--particle picture (band 
structure) and subsequently including the interaction via a \emph{local} 
potential, may not be appropriate. 
In such situations, one resorts to parametrized models of correlated 
electrons, 
where the single--particle and the interaction-induced aspects of the 
electronic states are treated on equal footing. 
The single--particle wave--functions are contained in the formal expressions 
of the model parameters. 
We have proposed \cite{spary} to combine the two efforts in an exact manner, 
at least for small systems.

In our method of approach (EDABI), we \emph{first} rigorously determine the 
ground--state energy $E_G$ of the system of interacting particles
using the occupation--number representation, which is expressed as a function 
of the microscopic parameters. 
\emph{Second}, we optimize this energy with respect to the wave--functions 
contained in 
these parameters by deriving the \emph{self--adjusted wave equation} for them.
Physically, the last step amounts to allowing the single--particle
wave functions to relax in the correlated state.
In practice, we propose the particular class of those functions, which are 
obtained by minimizing variationally the ground--state energy $E_G$
with respect to their size (effective Bohr radius).
In brief, our method of solution does not limit itself to an exact 
diagonalization of the paramerized Hamiltonian, but also involves an adjustment
of the single--particle wave function to obtain a true ground state of a
correlated quantum many--body system.

The EDABI method has been overviewed in a number of papers 
\cite{spacta,rylec,spacon}, so here we concentrate on its application to 
one--dimensional (1D) nanochains of $N\leqslant 16$ atoms, close to the 
metal--insulator crossover transition.
This paper complements our recent study of such systems \cite{spacta,rylec}
with the systematic analysis of both \emph{half--} and \emph{quarter--filled} 
band cases, as well as with the analysis of its transport properties.
Throughout the paper we use the adjustable Wannier functions composed of 
a Gaussian basis set (STO--3G), which are determined explicitly from
minimization of the system ground--state energy $E_G$ as a function of the 
interatomic distance $R$.

The question of delocalization of atomic states also has practical relevance.
Namely, in dealing with electronic properties of quantum dots one usually 
\emph{assumes} the existence of the effective--mass states. 
Whether this assumption is well founded for particular systems can only
be determined by finding the critical interatomic distance, above which 
the states are localized. 
In this context, we find such critical distance for model systems composed of 
$s$--like states.

The structure of the paper is as follows. In Section~\ref{edabi:intro} 
we briefly present the basic idea of the EDABI method.
Then, in Section~\ref{nchain}, we provide the numerical ground--state 
analysis of nanochains with the long--range Cou\-lomb interaction. 
Namely, we first discuss the electron momentum statistical distribution 
function for the half--filled case and analyse it in terms of 
Tomonaga--Luttinger scaling (including the logarithmic correction, 
\emph{cf.}\ Section \ref{tlmscal}) and
illustrate the charge--density wave ordering for the quarter filling
(\emph{cf.}\ Section \ref{onsetqf}), as well as the analysis the 
$N$--particle wave function localization according to the
method developed by Resta \cite{resta}.
In Section~\ref{gadrud}, we calculate the system charge and spin gaps, 
as well as perform the finite--size scaling with $1/N\rightarrow 0$ on these
quantities. 
We also analyze the spectral density and extract from it
the \emph{renormalized} dispersion relation for the interacting electrons 
in a nanochain. 
To the best of our knowledge, such an exact renormalized one--electron 
\emph{band structure} has not been determined before.
Finally, the application of the present scheme to the higher ($ns$--like) 
valence orbital systems would make possible a direct comparison with the 
experimental results for quantum nanowires made of noble and alkaline elements.
Also, the correspondence between the localization criteria infered from the 
ground--state properties on one side, and from the dynamical 
correlation functions on the other, is established.

\section{The combined exact diagonalization ab--initio (EDABI) method}
\label{edabi:intro}
 
\begin{figure}[!t]
\includegraphics[width=\columnwidth]{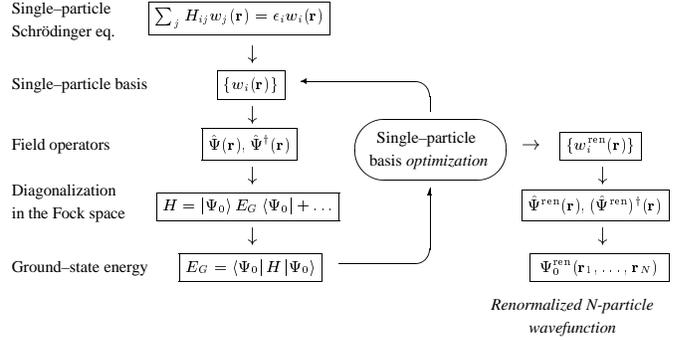}
\caption{Flowchart diagram of the EDABI method. 
The part on the far right provides \emph{renormalized} Wannier 
functions, field operators, and $N$-particle wave function.
The top line is absent if we select the trial Wannier functions as composed 
of adjustable Gaussian orbitals (e.g.\ the STO--3G basis).}
\label{edabi}
\end{figure}

The basic idea of the EDABI method \cite{spary,spacon} is illustrated on the 
block diagram exhibited in 
Fig.\ \ref{edabi}. We start from choosing the initial Wannier basis set 
$\{w_i({\bf r})\}$, composed of atomic--like wave functions (here a trial 
Gaussian basis set) with radii $a_i$. 
Next, we write down the system Hamiltonian in the second--quantization form 
and determine the ground--state energy $E_G$ together with the corresponding 
state in the Fock space $\ket{\Psi_0}$ by employing the Lanczos procedure, 
for example. 
Than, the Wannier basis set $\{w_i({\bf r})\}$ is optimized with respect to 
the atomic radii $a_i$, contained in the atomic wave functions (here 
represented by Gaussians) composing 
$w_i({\bf r})$, until the minimal ground--state energy $E_G$ is reached 
\cite{spary} for a given lattice parameter $R$. 
On the right, we list the renormalized quantities (for the optimized basis),
which can be calculated once the whole two--step procedure has provided
convergent results.

\section{The correlated electrons in a nanochain}
\label{nchain}
We consider the system of $N_e$ electrons on $N$ lattice sites arranged 
periodically, each site containing a single valence orbital and an 
infinite--mass ion (i.e.\ we start from hydrogenic--like atoms). 
The Hamiltonian, including \emph{all} the direct Cou\-lomb--in\-ter\-ac\-tion 
terms and neglecting other (e.g.\ exchange terms), can be written down 
(up to a constant) in the form
$$
  H=\ep_a^\eff\sum_i n_i 
  + t\sum_{i\sigma}\left(a_{i\sigma}^{\dagger}a_{i+1\sigma}+\mbox{HC}\right)
$$
\begin{equation}
\label{hameff}
  + U\sum_in_{i\ua}n_{i\da} + \sum_{i<j}K_{ij}{\de n_i}{\de n_j}, 
\end{equation}
where $\delta n_i\equiv n_i-1$, $\epsilon_a^{\rm eff}=
\epsilon_a+N^{-1}\sum_{i<j}(2/R_{ij}+K_{ij})$ (in Ry) is the effective atomic 
level, $R_{ij}$ is the distance between the $i$--th and $j$--th atoms, 
$t\equiv\bra{w_i}T\ket{w_{i+1}}$ is the nearest--neighbor hopping, 
$U\equiv\bra{w_iw_i}V\ket{w_iw_i}$ and $K_{ij}\equiv\bra{w_iw_j}V\ket{w_iw_j}$ 
are the intra-- and inter--site Cou\-lomb repulsions, respectively.
In the present form of the Hamiltonian, all the \emph{mean--field} Coulomb 
terms are collected in $\ep_a^\eff$, whereas the last term represents the 
\emph{correlation part} of the long--range Coulomb interaction.
We shall test \emph{a~posteriori} whether the tight--binding approximation for
the hopping term is valid. Also, the effect of the direct (Heisenberg) exchange
is negligible, since the kinetic exchange term will always be dominant 
\cite{rylec}. 
The microscopic parameters are expressed in terms of the Wannier functions
$\{w_i(\mathbf{r})\}$ composed of Gaussian--type orbitals.

The Hamiltonian (\ref{hameff}) is diagonalized in the Fock space with 
the help of Lanczos technique. As the microscopic parameters 
$\epsilon_a^{\rm eff}$, $t$, $U$, and $K_{ij}$ are calculated numerically
in the Gaussian basis, the orbital size of the 1$s$--like state expressed in
this basis is subsequently adjusted to obtain the minimal ground--state energy 
$E_G$, as a function of the interatomic distance $R$. 
Earlier, we have shown \cite{spary} that such a combined exact diagonalization
-- \emph{ab initio} study of the one dimensional system provides the 
localization threshold, the electron--lattice couplings, and the dimerization 
magnitude. 
Moreover, the utilization of the Gaussian--type orbitals leads to a variational
procedure that converges rapidly with the lattice size $N$ 
\cite{rylec,rycacta}. 
In effect, one can extrapolate the optimal orbital parameters for
larger $N$ using those obtained for small systems (i.e.\ for $N=6\div 10$), 
which speeds up the computation remarkably.
The purpose of this paper is to discuss basic solid--state properties of the
nanoscopic systems \emph{per se}, and (in some instances) their infinite
correspondants by performing the finite--size scaling.

\subsection{Statistical distribution and the To\-mo\-na\-ga--Luttinger 
scaling: The half--filled case}
\label{tlmscal}
We now discuss the electron momentum distribution of 1D chain of $N=6\div 16$ 
atoms to address the question of whe\-ther the system composes either 
a \emph{Luttinger--liquid} or forms an insulating (Mott--Hubbard) state.
We first summarize, following Voit \cite{sovo}, the properties
of 1D conductors, which include the two principal characteristics:
\begin{enumerate}
\item[$(i)$] 
A continuous momentum distribution function, showing a singularity near
the Fermi level $k\rightarrow k_F$ of the form (Solyom, Ref.\ \cite{sovo})
\begin{equation}
\label{llnkfun}
  n_{k\sigma}=n_F+A\abs{k_F-k}^{\theta}\sgn(k_F-k),
\end{equation}
where $\theta$ is a non--universal (\emph{interaction--dependent}) exponent; 
in con\-se\-quen\-ce, it leads to the absence of fer\-mio\-nic 
quasi--particles (the quasi--particle residue in vanishes as 
$z_k\sim\abs{k_F-k}^{\theta}$ when $k\rightarrow k_F$).
In other words, the \emph{Fermi ridge} is absent in this case.
\item[$(ii)$] 
Similar power--law behavior of all the other physical properties, particularly 
of the single--particle density of states, 
${\cal N}(\om)\sim\abs{\om-\mu}^{\theta}$ (i.e.\ a presence of a 
\emph{pseudogap}), that implies a Drude weight $D>0$ for $\theta<1$.
\end{enumerate}

In the case of lattice models, such as the (\emph{extended}) Hubbard model, 
the Luttinger liquid behavior is predicted by the renormalization group (RG) 
mapping onto the To\-mo\-na\-ga--Luttinger model \cite{sovo}.
Through such mapping, one can also expect, with the increasing $N$, a 
gradual convergence of the discrete momentum distribution $n_{k\sigma}$ into 
the continuous power--law form (\ref{llnkfun}).
This hypothesis was first checked numerically  for the Hubbard model 
by Sorella \emph{et al.} \cite{sore}.

\begin{table}[!t]
\caption{Microscopic parameters (in Ry) of Hamltonian (\ref{hameff}) for 
a nanochain calculated in the adjusted STO--3G basis composing the Wannier
functions. The numerical extrapolation with $N\rightarrow\infty$ is performed.
The values of the inverse orbital size $\alpha_\mathrm{min}$
(in the units of the Bohr radius $a_0$) and of the ground--state energy 
$E_G/N$ (for $N=10$), are also provided.}
\label{tuk:tab}
\begin{tabular}{rclllll}
 \hline\hline
 $R/a_0$ & $\alpha_\mathrm{min}a_0$ & $\epsilon_a^\mathrm{eff}$
 & $t$ & $U$  & $K_1$ & $E_G/N$ \\ \hline
 1.5 & 1.363 &  0.100 & -0.831 & 2.054 & 1.165 & -0.749 \\
 2.0 & 1.220 & -0.550 & -0.442 & 1.733 & 0.911 & -0.930 \\
 2.5 & 1.122 & -0.797 & -0.264 & 1.531 & 0.750 & -0.979 \\
 3.0 & 1.062 & -0.902 & -0.171 & 1.407 & 0.639 & -0.991 \\
% 3.5 \\
 4.0 & 1.013 & -0.971 & -0.080 & 1.291 & 0.493 & -0.992 \\
% 4.5 \\
 5.0 & 1.004 & -0.987 & -0.037 & 1.258 & 0.399 & -0.992 \\
 \hline\hline
\end{tabular}
\end{table}

Here we present the approach to a finite 1D chain with a long--range 
Cou\-lomb interaction, as described by the Hamiltonian (\ref{hameff}),
with a simultaneous evaluation of the model parameters by optimizing the
single--particle wave functions contained in those parameters \cite{spary}.
The values of the parameters in the ground state are shown in Table 
\ref{tuk:tab} as a function of the interatomic spacing (all quantities are in
atomic units). Only the value $K_1$ of the intersite Coulomb interaction 
$K_{ij}\equiv K_{\abs{i-j}}$ for nearest neighbors is listed, since more 
distant interactions scale essentially in the same manner as their classical
values, $K_{ij}\approx 2/\abs{\mathbf{R}_i-\mathbf{R}_j}$.
One should also note that the orbital size $\alpha^{-1}$ renormalized by the
electron--electron interaction is about $30-40\%$ smaller in the correlated
state than the corresponding value in the atomic limit ($\alpha^{-1}=a_0$).
Therefore, the tight--binding approximation made for the hopping term in Eq.\
(\ref{hameff}) is applicable to a good accuracy even for the lattice constant
$R/a_0=2.0$, but not much below this value.

\begin{figure}[!t]
\setlength{\unitlength}{0.01\columnwidth}
\begin{picture}(100,100)
\put(-36,-16){\includegraphics[width=1.75\columnwidth]{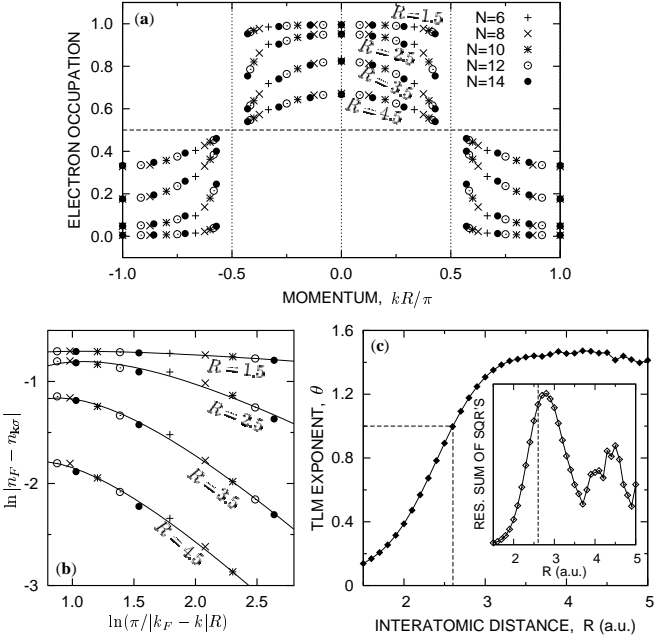}}
\end{picture}
\caption{Statistical distribution of electrons and Luttinger--liquid 
scaling for a \emph{half--filled} 1D chain of 
$N=6\div 14$ atoms with \emph{long--range} Coulomb interactions: 
$(a)$ momentum distribution for electrons in the linear and $(b)$ log--log 
scale; 
$(c)$ Tomonaga--Luttinger model exponent $\theta$ vs.\ lattice parameter $R$
(specified in $a_0$) and (\emph{inset}) the corresponding residual sum 
of squares.
Solid lines in Figs.\ $(a)$ and $(b)$ represent the fitting of Eq.\
(\ref{loloexp}).}
\label{nksll}
\end{figure}

The discrete electron--momentum distribution for the half--filling 
($N_e=N$) is depicted in Fig.\ \ref{nksll}a, while in Fig.\ \ref{nksll}b 
we replot it on a \emph{log--log} scale together with the fitted theoretical
curves, as explained below.
In order to obtain these results, we use the boundary conditions (BC) 
that minimize the ground--state energy for 
a given $N$ (namely, the \emph{periodic} BC for $N=4n+2$ atoms and the 
\emph{antiperiodic} BC for $N=4n$, at the half--filling).
A systematic arrangement of the distribution--function data in 
Fig.\ \ref{nksll}a for different $N$ values is striking, in that
for smaller $R$ it is \emph{Fermi--like}).

To extract the Luttinger--liquid exponent $\theta$ accurately from the data for
finite $N$, it was necessary to also include the higher scaling corrections to 
(\ref{llnkfun}). 
They can be obtained from the Tomonaga mapping in the form of an expansion
in the powers of $\ln(\pi/|k_F-k|R)$, namely,
\begin{equation}
  \label{loloexp}
  \ln\abs{n_F-n_{k\sigma}}=-\theta\ln z + b\ln\ln z + c
  + {\cal O}(1/\ln z),
\end{equation}
where $z\equiv\pi/|k_F-k|R$. 
This singular form of the expansion is required by the especially slow 
approach to the RG fixed point (Solyom, Ref.\ \cite{sovo}).
Obviously, by neglecting the logarithmic corrections one reaches the 
asymptotic form (\ref{llnkfun}) for $k\approx k_F$. 
The solid lines in Fig.\ \ref{nksll}b represent the formula 
(\ref{loloexp}); the fitted values of the parameters $\theta$, $b$, and $c$
are also listed in Table \ref{nksll:tab}.
The quality of the fit is decesively worse for points far away from 
the Fermi momentum, and depends on $N$ since the Fermi wave vector is
$N$--dependent, i.e.\ $k_F^N=k_F^{\infty}(1-2/N)$, where 
$k_F^{\infty}=\pi/(2R)$ represents the Fermi wave vector in the
$N\rightarrow\infty$ limit.
The exponent $\theta$ is also plotted in Fig.\ \ref{nksll}c as a function of
the lattice parameter $R$ showing that it crosses the critical value 
$\theta=1$ (corresponding to the metal--insulator boundary in 1D) for 
$R_\crit=2.60 a_0$ ($a_0=0.529\,\mbox{\AA}$ is the Bohr radius). 
We also give the \emph{residual sum of squares} 
(\emph{cf.}\ inset in Fig.\ \ref{nksll}c), which shows that the quality of the 
fit becomes worst for $R\approx R_\crit$ where the system approaches the 
localization threshold.

The results for the half--filled system with the \emph{on--site} Hubbard 
repulsion alone, and with the atomic energy part included explicitly as a 
function of $R$ \cite{thesis}, are qualitatively very similar to those 
displayed in Fig.\ \ref{nksll}. 
The critical value of the lattice parameter in this case is $R_\crit=2.16 a_0$,
and does not differ drastically from the previous one (\emph{cf.}\ Table 
\ref{nksll:hubb:tab} for the corresponding values of all the fitted 
parameters in the Hubbard--model case). 
This is because such nanoscopic systems always have a finite conductivity
in the large--density limit, since electrons tunnel through a finite--width
and finite--height potential barrier. 
Therefore, such half--filled systems, both with and without inclusion 
of the long--range interactions, can be considered as being close to the 
metal--insulator transition, in no apparent contradiction with the 
infinite--chain RG result by Fabrizio \cite{fabri}, and the 
Hubbard--model solution by Lieb and Wu \cite{liebwu}, which both provide only
the Mott insulating behavior.
This discussion is complete only after calculating the charge and spin gaps, 
as well as the electric conductivity, which are dealt with in the next two 
Sections.

\begin{table}[!t]
\caption{The parameters of the expansion (\ref{loloexp}) for the 
\emph{half--filled} chain with \emph{long--range} Coulomb interactions. 
The corresponding standard deviations $\sigma(X)$ for the quantities 
$X=\theta, b$ and $c$ are also specified.}
\label{nksll:tab}
\begin{tabular}{rllllll}
 \hline\hline
 $R/a_0$ & $\theta$ & $\sigma(\theta)$
 & $b$ & $\sigma(b)$  & $c$ & $\sigma(c)$ \\ \hline
 1.5  & 0.138 & 0.015 & 0.147 & 0.024 & -0.567 & 0.015 \\
 2.0  & 0.387 & 0.055 & 0.425 & 0.089 & -0.346 & 0.053 \\
 2.5  & 0.893 & 0.122 & 0.971 & 0.196 & $ $ 0.084 & 0.118 \\
 3.0  & 1.307 & 0.128 & 1.315 & 0.207 & $ $ 0.357 & 0.125 \\
% 3.5  & 1.433 & 0.085 & 1.264 & 0.137 & $ $ 0.262 & 0.082 \\
 4.0  & 1.455 & 0.186 & 1.113 & 0.299 & -0.032 & 0.180 \\
% 4.5  & 1.462 & 0.109 & 1.057 & 0.176 & -0.384 & 0.106 \\
 5.0  & 1.413 & 0.133 & 0.943 & 0.214 & -0.823 & 0.129 \\
 \hline\hline
\end{tabular}
\end{table}

\begin{table}[!t]
\caption{The fitted parameters of the singular expansion (\ref{loloexp})
  for a half--filled 1D Hubbard chain.}
\label{nksll:hubb:tab}
\begin{tabular}{r|ll|ll|ll}
 \hline\hline
 $R/a_0$ & $\theta$ & $\sigma(\theta)$
 & $b$ & $\sigma(b)$  & $c$ & $\sigma(c)$ \\ \hline
 1.5  & 0.229 & 0.030 & 0.237 & 0.048 & -0.537 & 0.029 \\
 2.0  & 0.803 & 0.100 & 0.855 & 0.162 & -0.078 & 0.097 \\
 2.5  & 1.283 & 0.109 & 1.259 & 0.176 & $ $ 0.217 & 0.106 \\
 3.0  & 1.420 & 0.075 & 1.230 & 0.121 & $ $ 0.116 & 0.073 \\
% 3.5  & 1.415 & 0.048 & 1.083 & 0.078 & -0.163 & 0.047 \\
 4.0  & 1.436 & 0.069 & 1.033 & 0.111 & -0.456 & 0.067 \\
% 4.5  & 1.445 & 0.166 & 1.000 & 0.268 & -0.801 & 0.161 \\
 5.0  & 1.371 & 0.037 & 0.873 & 0.060 & -1.218 & 0.036 \\
 \hline\hline
\end{tabular}
\end{table}

The present analysis supplements the earlier discussion 
\cite{spary,rylec,rycacta}, 
in which we have interpreted the distribution $n_{\mathbf{k}\sigma}$
in Fig.\ \ref{nksll}a in terms of the modified Fermi distribution for an 
\emph{almost localized} Fermi liquid.
The points are arranged in an almost flat manner for the interatomic distance
$R=1.5a_0$, suggesting that some kind of quasi--discontinuity of 
$n_{\mathbf{k}\sigma}$ exists near $k_F$.
However, an ambiguity arises because of the circumstance that for 
nanosystems, the points very close to the Fermi momentum for 
$N\rightarrow\infty$ system, i.e.\ the point $k_F^{\infty}=\pi/(2R)$, are 
simply missing.
Nonetheless, it is amazing that the momentum distribution can be rationalized 
in such simple terms (as either the Fermi or the Tomonaga--Luttinger liquids), 
which represent concepts borrowed from the $N\rightarrow\infty$ limit.
One should also say that the critical value of $R_\crit=2.60a_0$ for the
localization of the single--particle states obtained here is about $30\%$
lower than the corresponding value $R_\crit=3.93a_0$ obtained when we treat
the distribution $n_{\mathbf{k}\sigma}$ as the modified Fermi distribution.
However, a discreapancy of this order should not be suprising anybody, since 
the localization criteria should be treated as semi--quantitive at best. 
One can hope to clarify the situation by extending the present analysis to
larger $N$.
Nevertheless, if we regard a nanoscopic system of $N\sim 10$ atoms as 
a \emph{real systems}, then the ambiguity of the statistical
distribution is significant.
Also, the \emph{partial localization} of electrons in a nanosystem will 
become apparent when we discuss the multiparticle wave--function localization 
at the end of this Section, and the conductivity in the \emph{next} Section.

We allow ourselves one specific suggestion (if not speculation) at this 
point. Namely, the results shown in Fig.\ \ref{nksll} (and the discussion
in the remaining part of the paper) point to the possibility that the system
with small $R\lesssim 2a_0$ can be analyzed as a
Landau--Fermi liquid (albeit with discrete momentum states \cite{altshu}),
whereas the system with $2a_0\lesssim R\lesssim 2.5a_0$ is closer to the
Tomonaga--Luttinger--Solyom limit. For $R\gtrsim 2.5a_0$ the electrons can be
regarded as effectively localized. Such a division into \emph{three physically 
distinct} regimes requires a further discussion, carried out, to some extent, 
below.

Finally, it should be noted that a well--defined spin--spin correlations
of the \emph{spin--density--wave} type develop with the increasing $R$ for 
the half--filled case, as discussed elsewhere \cite{rylec,thesis}.
The system ground--state energy as a function of $R$ is also provided there.

\subsection{Onset of the charge--density wave state for the quarter filling}
\label{onsetqf}

\begin{figure}[!b]
\setlength{\unitlength}{0.01\columnwidth}
\begin{picture}(100,60)
\put(-25,-22){\includegraphics[width=1.6\columnwidth]{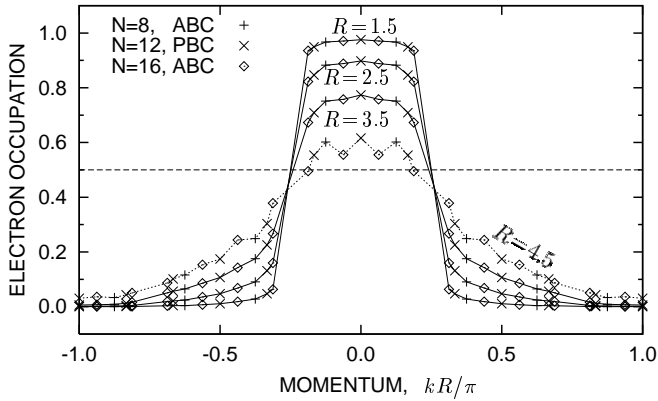}}
\end{picture}
\caption{Momentum distribution $n_{k\sigma}$ for electrons on a chain of
$N=8\div 16$ atoms in the \emph{quarter--filled} case ($N_e=N/2$). 
Lines are drawn as a guide to the eye only.
ABC and PBC denote the \emph{antiperiodic} and \emph{periodic} boundary 
conditions.}
\label{nksqf}
\end{figure}

\begin{figure}[!b]
\setlength{\unitlength}{0.01\columnwidth}
\begin{picture}(100,50)
\put(-33,-16){\includegraphics[width=1.75\columnwidth]{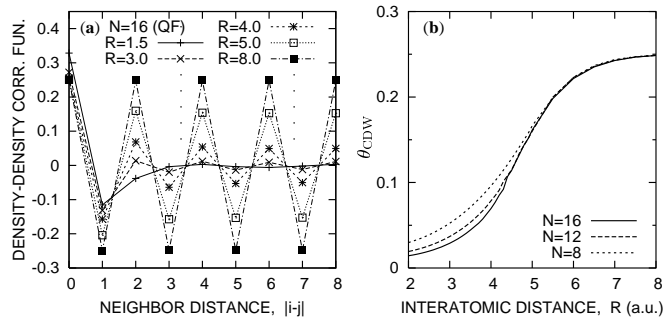}}
\end{picture}
\caption{Charge--density distribution for the \emph{quarter--filled} 
($N_e=N/2$) nanochain: 
$(a)$ density fluctuation correlation function 
$\aver{\Delta n_i\Delta n_j}$ vs.\ distance $\abs{i-j}$, 
$(b)$ \emph{charge--density wave} order parameter for the  
density--density fluctuation (\emph{see} main text for the definition) 
vs.\ interatomic distance $R$.}
\label{corfqf}
\end{figure}

The electron quasi--momentum distribution for the \emph{quar\-ter--filled} 
(QF) case ($N_e=N/2$) is shown in Fig.\ \ref{nksqf}. 
The available number of data points is too small to fit the singular formula
(\ref{loloexp}) to a reasonable accuracy. In effect, the lines on the plots 
are a guide for the eye. 
However, the smooth behavior of the Luttinger--liquid type is evident for $R
\lesssim 4a_0$, and changes dramatically  for the larger values of $R$. 
This change reflects the onset of the \emph{charge--density--wave ordering}, 
as illustrated in Fig.\ \ref{corfqf}.
In a QF chain of $N=16$ atoms (\emph{cf.}\ Fig.\ \ref{corfqf}a) the charge is 
almost uniformly distributed for $R\lesssim 3a_0$, but the charge density wave
(CDW) sets in very rapidly in the range $R/a_0=4\div 5$. 
The CDW order parameter, defined as 
$
  \theta_{\rm CDW}\equiv N^{-1}\sum_m(-1)^m\aver{\Delta n_i\Delta n_{i+m}},
$
(where $\Delta n_i\equiv n_i\!-\!\averaa{n}$) approaches its maximal value 
$\theta_{\rm CDW}=1/4$ for $R\gtrsim 8a_0$ (\emph{cf.}\ Fig.\ \ref{corfqf}b).
Also, the crossover range of $R$, where $\theta_{\rm CDW}$ evolves from 
$\theta_{\rm CDW}\approx 0$ to the perfect--order value $\theta_{\rm CDW}=1/4$,
shrinks systematically with the increasing $N$, suggesting \emph{quantum 
critical behavior} in the large--$N$ limit. 
One can argue that the charge--ordered state for larger $R$, corresponding
to 1D \emph{Wigner--crystal} state on a lattice, is unstable in the 
$N\rightarrow\infty$ limit, in accordance with the Mermin--Wagner theorem. 
Namely, we expect the amplitude of the quantum charge fluctuations to diverge 
at zero temperature as $\log N$ for
the system with a long--range ($\sim 1/r$) Coulomb coupling.
However, the divergence is absent for $R\gtrsim 5a_0$ in the 
exact--diagonalization data for $N=8\div 16$ (\emph{cf.}\ Fig.\ \ref{corfqf}b),
indicating that the zero--point charge fluctuations are supressed by the 
onset of the CDW state.
In other words, the larger lattice parameter $R$, the larger the system
which can be regarded as charge ordered in the quarter--filled case. 
This notion agrees with the expected atomic--limit charge--order in the QF 
case.
The correspondence between the appearance of such charge order and the 
system conductivity is discussed in the \emph{next} Section, but first,
we supplement our analysis of the system ground--state properties with 
the many--body wavefunction localization in the framework proposed by Resta 
\cite{resta}.

\subsection{Ground--state wavefunction localization}
The approach \cite{resta} to the electron localization in the correlated
state is based on the idea that, although the insulating or the metallic 
states of matter are usually characterized by their excitation spectrum, 
the qualitative difference in \emph{dc} conductivity must also reflect a 
qualitative difference in the organization of electrons in their ground state. 
Such a concept was first emphasized by Kohn in a milestone paper \cite{kohn}, 
but a complete treatment, related to the Berry--phase theory of
polarization, was proposed over 30 years later by Resta and Sorella 
(\emph{cf.}\ Ref.\ \cite{reskor}). 
In their approach the complex number
\begin{equation}
\label{zndef}
  z_{N_e}=\bra{\Psi_0}e^{i(2\pi/L)\hat{X}}\ket{\Psi_0}
\end{equation}
(where $\hat{X}$ is a many--body position operator defined in 1D as 
$\hat{X}\equiv\sum_jx_j$ and $L=NR$ is the system length) is used to 
discriminate between a localized  $N_e$--particle ground state 
(where $\abs{z_{N_e}}\rightarrow 1$ for large 
systems) and a delocalized one, where $z_{N_e}$ vanishes.
Namely, the qualitative measure of the electron localization is defined as
\begin{equation}
\label{x2cdef}
  \aver{x^2}_c=
  -\frac{1}{N_e}\left(\frac{L}{2\pi}\right)^2\log\abs{z_N}^2=
  -\frac{NR^2}{4\pi^2\bar{n}}\log\abs{z_N}^2
\end{equation}
(where $\bar{n}=N_e/N$ is the average electron density), whereas the phase 
of the complex number $z_{N_e}$ is related to the macroscopic system 
polarization.
One should note at this point that the kind of localization described by Eq.\
(\ref{x2cdef}) is clearly \emph{not} a feature characterizing the individual 
single--particle orbitals; instead, it is a global property of a 
multiparticle ground--state wavefunction \emph{as a whole}. 
That is because the operator $e^{i(2\pi/L)\hat{X}}$
in Eq.\ (\ref{zndef}) \emph{cannot} be expresed as a single--particle operator
(like $\hat{X}$) but rather as a \emph{genuine} many--body operator.

\begin{figure}[!t]
\setlength{\unitlength}{0.01\columnwidth}
\begin{picture}(100,118)
\put(-25,-20){\includegraphics[width=1.6\columnwidth]{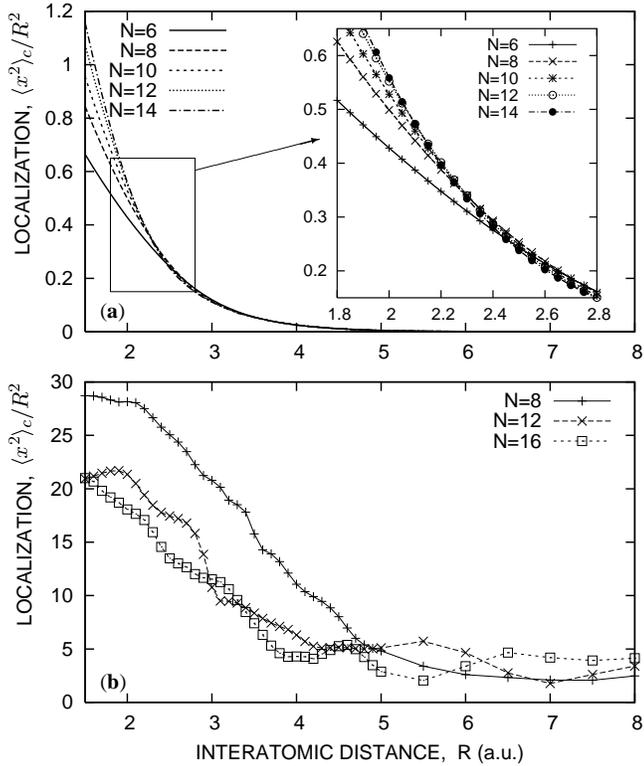}}
\end{picture}
\caption{Localization parameter for the ground--state multiparticle 
wavefunction of the \emph{half--filled} $(a)$ and the \emph{quarter--filled} 
$(b)$ nanochains.
The values of the localization parameter (\emph{see} main text for the 
definition) are specified in the units of $R$.}
\label{x2loc}
\end{figure}

The numerical evaluation of the complex number $z_{N_e}$ and the resulting
$\averaa{x^2}_c$
within the Lanczos algorithm is straightforward, since the operator 
$e^{i(2\pi/L)\hat{X}}$ in Eq.\ (\ref{zndef}) is diagonal in the position 
representation.
The results for a 1D system with long--range Coulomb interaction are shown
in Fig.\ \ref{x2loc} for both the half-- and the quarter--filled cases 
(\emph{cf.}\ Fig.\ \ref{x2loc}a and \ref{x2loc}b, respectively).
To avoid confusion by changing the system length $L$, we plot the
localization data in units of the lattice parameter $R$, namely 
$\averaa{x^2}_c/R^2$, instead of $\averaa{x^2}_c$. 
The values of the localization parameter, depicted in Fig.\ \ref{x2loc}a,
gradually decay with growing $R$ for all available numbers of atoms $N$. 
However, the dramatic change of the decay character takes place around the 
value of the lattice parameter $R\approx 2.5a_0$ (cf.\ \emph{inset} in Fig.\ 
\ref{x2loc}a), where the curves for different $N$ join together and form a 
single one for $R\gtrsim 2.5a_0$. In contrast, for lower values of $R$, the 
curves are well separated there by showing the divergence of the localization 
parameter with $N$ for $R\lesssim 2a_0$. 
The results for the Hubbard model (not shown) again does not differ 
qualitatively from those presented in Fig.\ \ref{x2loc}a; the coalescence 
point is located near the value $R\approx 2.2a_0$. 
The positions of the joining points for both the models (with and without
inclusion of the long--range Coulomb interactions) are very close to the 
corresponding critical values $R_c$ obtained form the Tomonaga--Luttinger 
scaling at the begining of this Section.

The numerical results for the quarter--filled case, shown in Fig.\ 
\ref{x2loc}b, are of lesser accuracy than those for the half--filling. 
This is because the complex expectation value $z_{N_e}$, defined by Eq.\ 
(\ref{zndef}), is itself a sum of many complex numbers with different phases,
when calculated in the postition representation. 
If the resulting $z_{N_e}$ is close to zero, which is the case for QF for
all examined values of $R$; it is, in turn, strongly affected by the computer
roundoff errors, which are also amplified by taking the logarithm of
$\abs{z_{N_e}}^2$ when calculating $\averaa{x^2}_c$ from Eq.\ (\ref{x2cdef}).
Nevertheless, the evolution of the wavefunction localization parameter
versus $R$ is qualitatively similar to that for the half--filed case: the 
curves for different system sizes $N$, depicted in Fig.\ \ref{x2loc}b, get 
very close to each other for $R\gtrsim 4.5a_0$ where the charge--density wave
state is formed, as discussed in the preceding subsection.

The apparent correspondence between the wavefunction localization properties 
and the nature of electron momentum distribution, both discussed in this 
Section, suggests that a significant reorganization of the ground state takes 
place when the nanochain is close to the localization thre\-shold.
These observations are supplemented in the \emph{next} Section with analysis 
of the system energy gap, the spectral function, and the transport properties.

\section{Spectral and transport properties}
\label{gadrud}
\subsection{The charge and spin gaps}

\begin{figure}[!t]
\setlength{\unitlength}{0.01\columnwidth}
\begin{picture}(100,60)
\put(-25,-22){\includegraphics[width=1.6\columnwidth]{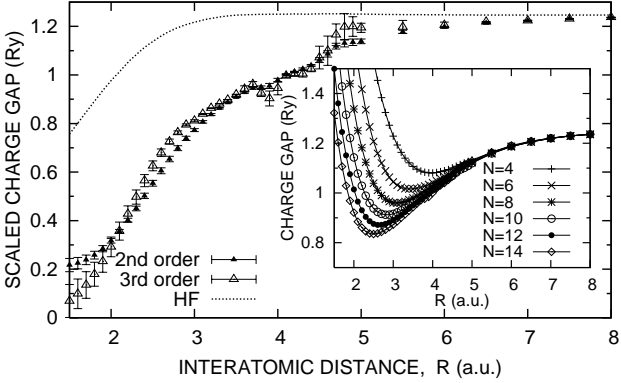}}
\end{picture}
\caption{Charge--energy gap obtained through the finite--size scaling of the 
results for the chains of $N=4\div 14$ atoms. 
The corresponding Hartree--Fock (HF) value of the magnetic gap
for an infinite system is also drawn for comparison (\emph{dotted} line). 
The \emph{inset} exhibits the original data for different values of $N$, 
used for the scaling. \emph{Note} a qualitative difference between the scaled
(with $1/N\rightarrow 0$) and the actual gap values.}
\label{delc}
\end{figure}

For a further verification, of whether the system is quasi--metallic or 
quasi--insulating in the Lut\-tin\-ger--liquid like regime, we first perform 
an extrapolation with $1/N\rightarrow 0$ of the charge--gap defined (for the 
\emph{half--filling}) as
\begin{equation}
\label{engap}
  \Delta E_C(N)=E_G^{N+1}(N)+E_G^{N-1}(N)-2E_G^{N}(N),
\end{equation}
where $E_G^{N_e}(N)$ is the ground-state energy of the $N$--site system 
containing $N_e$ electrons. 
The corresponding numerical results are shown in Fig.\ \ref{delc}, where we
use again the proper boundary conditions (\emph{periodic} or 
\emph{antiperiodic}, depending on $N$) which minimize the ground--state energy.
The extrapolation of $1/N\rightarrow 0$ performed using the $2$--nd and 
the $3$--rd order polynomials in $(1/N)$ provides a nonzero value of 
$\Delta E_C$ for wide range of the lattice parameter $R$. Only for the lowest 
examined value of $R=1.5a_0$ does the gap $\Delta E_C$ reaches zero within 
the extrapolation error; for $R>1.5a_0$ it is clearly nonvanishing. 
The gap is also significantly smaller than the corresponding Hartree--Fock 
value (\emph{dotted} line) in the regime $R\lesssim 4.5a_0$, suggesting that 
some kind of reorganization is present in the dielectric properties, e.g.\ a 
crossover from the Slater-- to the Mott--type insulator, as discussed for the 
parametrized models \cite{reskor}.
This hypothesis is verified by estimating the spin gap \emph{below}. 
One should also note a qualitative difference between the behavior of 
$\Delta E_C$ for nanoscopic chains (\emph{cf.}\ the inset) and of their 
$N\rightarrow\infty$ correspondants. Therefore, the finite--$N$ system 
contains physically different dynamic properties.

\begin{figure}[!t]
\setlength{\unitlength}{0.01\columnwidth}
\begin{picture}(100,60)
\put(-25,-23){\includegraphics[width=1.6\columnwidth]{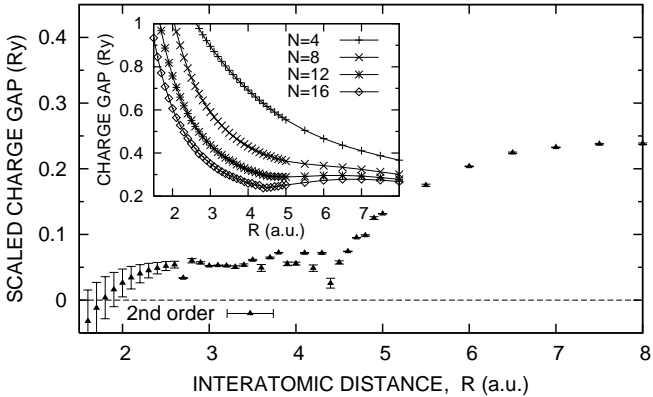}}
\end{picture}
\caption{Charge--energy gap obtained through the finite--size scaling of the
results for the chains of $N=4\div 16$ atoms in the \emph{quarter--filled} 
case. The $2$--nd order polynomial has been fitted to perform the 
extrapolation with $1/N\rightarrow 0$. \emph{Inset} provides the original data 
used for the scaling.}
\label{delc:qf}
\end{figure}

The situation changes when we consider the \emph{quarter--filled} case, 
$N_e=N/2$ (\emph{cf.}\ Fig.\ \ref{delc:qf}).
The pa\-ra\-bo\-lic extrapolation with $1/N\rightarrow 0$ now provides the 
value of the charge--gap $\Delta E_C\approx 0$ (within the error bars) for the 
lattice parameter $R\lesssim 2a_0$. 
In the range of $R/a_0=2.5\div 4.5$ the gap develops (it is significantly 
greater than the corresponding error bars), but a random dispersion of the 
data points suggests an inaccuracy of the extrapolation due to the nonanalytic
behavior of $\Delta E_C$ when the system approaches the localization 
threshold. 
For $R\gtrsim 4.5a_0$ the gap grows smoothly to a limiting value corresponding
to that for the insulating charge--density wave state, identified earlier.
The more precise position of the localization point is determined later in
this Section, where we calculate the system Drude weight.

\begin{figure}[!t]
\setlength{\unitlength}{0.01\columnwidth}
\begin{picture}(100,50)
\put(-34,-17){\includegraphics[width=1.75\columnwidth]{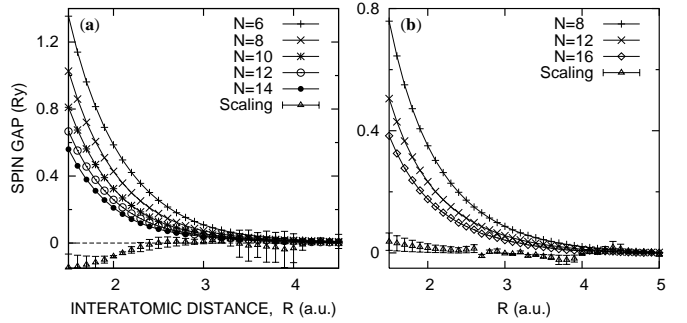}}
\end{picture}
\caption{Spin gap for the \emph{half--filled} $(a)$ and the 
\emph{quarter--filled} $(b)$ nanochains. 
The results obtained through the finite--size scaling are also shown with
the corresponding errorbars.}
\label{dels}
\end{figure}

For the sake of completness, in Fig.\ \ref{dels} we provide the values of 
the system \emph{spin gap}, which is defined as
\begin{equation}
\label{delsdef}
  \Delta E_S(N)\equiv E_G^{S_z=1}(N)-E_G^{S_z=0}(N), 
\end{equation}
where $E_G^{S_z}(N)$ denotes the lowest eigenenergy of $N$--site system in the 
subspace with a given total $z$--component of spin $S_z$.
Through the finite--size scaling with $1/N\rightarrow 0$, we obtain the spin 
gap $\Delta E_S\approx 0$ for both the half-- and the quarter--filled cases
(\emph{cf.}\ Figs.\ \ref{dels}a and \ref{dels}b, respectively), and in the 
whole examined range of the interatomic distance $R$. 
These results clearly show that the insulating phase of the system is of the
Mott type in the large--$N$ limit. 
No indications of the Slater--type phase, for which $\Delta E_C=\Delta_S>0$,
was found. 
However, for a nanosystem containing $N\sim 10$ atoms, one can note
that the so--called \emph{correlated insulator} ($\Delta E_C>\Delta_S>0$), 
existing for the small values of $R$ (cf.\ \emph{insets} to Figs.\ \ref{delc}
and \ref{delc:qf}, and Fig.\ \ref{dels}) gradually transforms into the Mott 
insulator with increasing $R$.
This evolution should be contrasted with the evolution of bulk 3D systems
where the Slater antiferromagnet evolves into the Mott insulator 
(\emph{cf.}\ Korbel \emph{et al.}, Ref.\ \cite{reskor}).

\subsection{Spectrum of single--particle excitations:
renormalized bands vs.\ Hubbard subbands}

\begin{figure}[!b]
\setlength{\unitlength}{0.01\columnwidth}
\begin{picture}(100,84)
\put(-10,-18){\includegraphics[width=1.45\columnwidth]{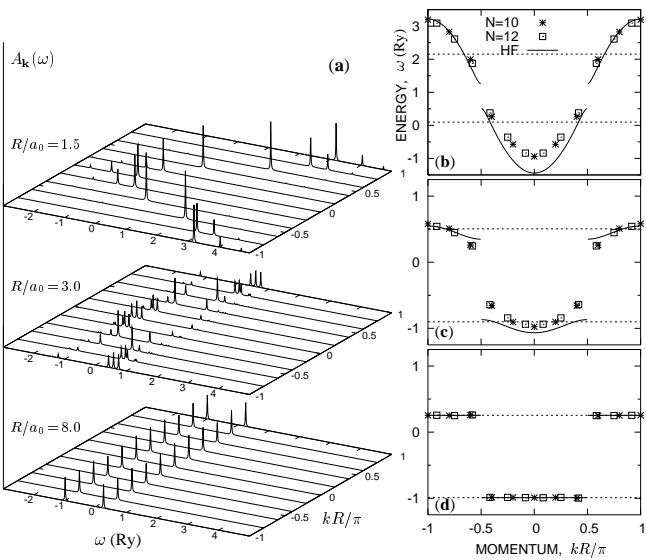}}
\end{picture}
\caption{Spectral functions $A_\mathbf{k}(\omega)$ for the nanochain of $N=10$
atoms $(a)$ and the corresponding renormalized dispersion relations (band 
energies with a gap) for the lattice parameter $R=1.5a_0$ $(b)$, 
$R=3a_0$ $(c)$, and $R=8a_0$ $(d)$. 
The quasiparticle energies for \emph{antiferromagnetic} Hartree--Fock solution
(\emph{solid} lines) are also shown for the comparison.}
\label{ak3d}
\end{figure}

The evolution of the single--particle spectral density function 
$A_\mathbf{k}(\omega)$ 
with increasing $R$ is shown in Fig.\ \ref{ak3d}a for the \emph{
half--filled} ($N_e=N$) nanochain described by the Hamiltonian (\ref{hameff}). 
The spectral function is defined in the standard manner, namely
\begin{equation}
\label{akw}
  A_\mathbf{k}(\omega)=\sum_n\abs{\bra{\Psi_n^{N\pm 1}}%
    c_{\mathbf{k}\sigma}^{\pm}\ket{\Psi_0^N}}^2
  \delta\left[\omega-\left(E_n^{N\pm 1}-E_0^N\right)\right],
\end{equation}
where the upper (lower) sign correspond to $\om>\mu$ ($\om<\mu$), 
respectively, $\ketaa{\Psi^N_n}$
is the $n$-th eigenstate of the system containing $N$ particles, $E_n^N$ is
the corresponding eigenenergy, and the matrix element
$\braaa{\Psi_n^{N\pm 1}}c_{{\bf k}\sigma}^{\pm}\ketaa{\Psi_0^N}$, with 
$c_{{\bf k}\sigma}^{+}\equiv a_{{\bf k}\sigma}^{\dagger}$ and
$c_{{\bf k}\sigma}^{-}\equiv a_{{\bf k}\sigma}$, is calculated within the 
Lanczos technique set up by Dagotto \cite{dago}.
For plotting purposes, we have used the analytical representation of the 
Dirac delta function
$\delta (x)\rightarrow (1/\pi) \ep/(x^2+\ep^2)$ with $\ep=0.01\ \mathrm{Ry}$;
this numerical trick leads to peaks with nonzero width, which otherwise would
be discrete.
In the nanometallic range ($R\lesssim 2.5a_0$), the quasiparticle peaks are 
well defined, but incoherent tails are always present and grow in strength 
with increasing $R$. 
In effect, in the intermediate regime of $R\sim 3a_0$ the lower and upper 
Hubbard bands are formed, which, in turn, continuously evolve into 
discrete atomic levels located at the positions $\om=\ep_a$ and $\om=\ep_a+U$,
when $R\rightarrow\infty$ (\emph{cf.}\ Fig.\ \ref{ak3d}d). 
These limiting peak positions correspond to the ground (H$^0$) and excited 
(H$^-$) atomic levels. 
Possibly the most interesting feature of this spectrum is its 
\emph{incoherent } nature for $R\sim 3a_0$, where the band and the interaction 
energies are comparable and where the Luttinger liquid exponent crosses the 
critical value $\theta=1$ (\emph{cf.}\ Section \ref{tlmscal}), corresponding 
to the localization threshold.

A direct picture of the spectrum evolution is provided by the 
\emph{renormalized} dispersion relation, which is ob\-tai\-ned from the 
spectral function by extracting the major quasiparticle peak position for each
momentum $k$, and is plotted in Figs.\ \ref{ak3d}b--d.
We can again easily identify: $(i)$ the \emph{na\-no\-me\-ta\-lic range} 
(\emph{cf.}\ Fig.\ \ref{ak3d}b for $R=1.5a_0$), for which the charge gap is 
of the same order as the energy discretization due to the geometrical 
quantization of the quasiparticle momenta, 
$(ii)$ the \emph{intermediate regime} (\emph{cf.}\ Fig.\ \ref{ak3d}c, $R=3a_0$)
where the gap becomes significantly wider, and $(iii)$ the \emph{atomic limit} 
(\emph{cf.}\ Fig.\ \ref{ak3d}d, $R=8a_0$) in which two dispersionless 
manifolds located at energies $\om=\ep_a$ and $\om=\ep_a+U$ appear. 
One should also note that the corresponding results for the antiferromagnetic
Hartree--Fock solution (\emph{solid} lines) closely match those obtained within
the EDABI method, both for the \emph{atomic} and for the \emph{nanometalic} 
limits (particularly for the upper Hubbard band in the latter case). 
The electron--correlation effects, in contrast, clearly appear (near the 
Fermi momenta) for the intermediate range of $R$, where the Hartree--Fock 
charge gap is significantly larger than the exact one (\emph{cf.}\ also the 
\emph{preceding} Section).
Another interesting feature of the renormalized dispersion relation is that
the da\-ta\-points for different $N$ ($=10$ and $12$) compose a single
(\emph{universal}) renormalized band--like dispersion relation 
$\tilde{\ep}_{\mathbf{k}}$, provided that the proper boundary conditions
which minimize the ground--state energy for a particular $N$ are applied.

The results for the renormalized band energy $\tilde{\ep}_{\mathbf{k}}$, 
together with those for the distribution function $n_{\mathbf{k}\sigma}$ 
(\emph{cf.}\ Fig.\ \ref{nksll}a), characterize in a \emph{fudamental manner} 
the salient features of the electronic states for the 
\emph{quantum nanoliquid}. 
Also note, that the gap in the single--particle spectrum is always present,
that is partially caused by the absence of the $k$ states near the Fermi point
$k_F^{\infty}$, and partially by the electron--electron interaction. 
The role of the geometrical momentum quantization is dominant in the 
large--density limit ($R\lesssim 2a_0$, \emph{cf.}\ Fig.\ \ref{ak3d}b), 
whereas the Coulomb repulsion determines the single--particle gap in the 
large--$R$ limit (\emph{cf.}\ Fig.\ \ref{ak3d}d). 
What is remarkable at this point, is that the Hartree--Fock aproximation works
well in both limits of weak and strong electron correlations mentioned
above, but fails in the intermediate range ($R\sim 3a_0$, \emph{cf.}\ 
Fig.\ \ref{ak3d}c).
One also can say that we observe a magnetic (Slater) gap contribution to 
the renormalized band structure of nanoscopic system even though we only 
observe SDW--like correlations, but \emph{no} spontaneous symmetry breaking.
This is another \emph{fundamental feature} of these systems.
The above results will now be supplemented with the transport properties, 
which are considered next.

\subsection{Drude weight and optical conductivity}
\label{chasti}

The real part of the optical conductivity at zero temperature is determined by 
the corresponding real part of the linear response to the applied electric 
field \cite{shamil}, and can be written as $\si(\om)=D\de(\om) + 
\si_{\rm reg}(\om)$, where the regular part is
\begin{equation}
\label{sireg:def}
  \si_{\rm reg}(\om)=\frac{\pi}{N}\sum_{n\neq 0}\frac{\abs{\bra{\Psi_n}j_p
  \ket{\Psi_0}}^2}{E_n-E_0}\de\left(\om-(E_n-E_0)\right),
\end{equation}
whereas the Drude weight (the \emph{charge stiffness}) $D$ is defined by
\begin{equation}
\label{d:def}
  D=-\frac{\pi}{N}\bra{\Psi_0}T\ket{\Psi_0}-\frac{2\pi}{N}
    \sum_{n\neq 0}\frac{\abs{\bra{\Psi_n}j_p\ket{\Psi_0}}^2}{E_n-E_0},
\end{equation}
where the kinetic--energy term $T$ is the same as the second term in Eq.\ 
(\ref{hameff}), and the diamagnetic current operator defined as usual:
$j_p= it\sum_{j\si}(a_{j\si}^{\dagger}a_{j+1\si}-\mbox{HC})$.
Here the states $\ket{\Psi_n}$ in Eqs.\ (\ref{sireg:def}) and (\ref{d:def})
are the eigenstates of the Hamiltonian (\ref{hameff}) corresponding to the
eigenenergies $E_n$, again with boundary conditions which minimize the 
ground--state energy for a given system size $N$. 
Matrix elements $\bra{\Psi_n}j_p\ket{\Psi_0}$ are calculated within the 
Lanczos method \cite{dago}. 
For plotting purposes,
we also again use the analytic representation of Dirac delta function
$\delta (x)\rightarrow (1/\pi) \ep/(x^2+\ep^2)$, with $\ep=0.01\ {\rm Ry}$.

For a finite system of $N$ atoms, $D$ is always nonzero due to a nonzero 
tunneling probability through a potential barrier of finite width.
Because of that, the finite--size scaling with $1/N\rightarrow 0$ 
has to be performed on $D$.
Here we use, after G\'{o}ra \emph{et al.} \cite{gora}, the following 
parabolic extrapolation
\begin{equation}
\label{lndnscal}
  \ln|D^*_N|=a+b(1/N)+c(1/N)^2,
\end{equation}
where $D^*_N$ denotes the \emph{normalized Drude weight} 
$D^*\equiv -(N/\pi)D/\bra{\Psi_0}T\ket{\Psi_0}$
for the system of $N$ sites, which provides the value in the range 
$0\leqslant D^*\leqslant 1$, and thus can be regarded as an alternative order 
parameter for the transition to the localized (atomic) state.

\begin{figure}[!t]
\setlength{\unitlength}{0.01\columnwidth}
\begin{picture}(100,50)
\put(-34,-17){\includegraphics[width=1.75\columnwidth]{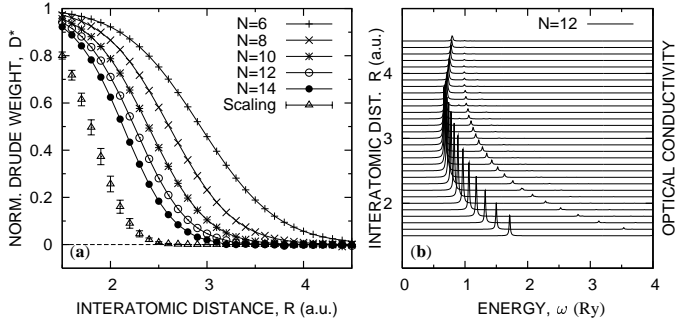}}
\end{picture}
\caption{Optical conductivity for nanochains in the \emph{half--filled} case: 
$(a)$ normalized Drude weight vs.\ lattice parameter $R$ (specified in $a_0$) 
and its values obtained through the finite size scaling ($1/N\rightarrow 0$); 
$(b)$ regular part of the conductivity, $\si_{\rm reg}(\om)$ for $N=12$ atoms.}
\label{sighalf}
\end{figure}

\begin{figure}[!t]
\setlength{\unitlength}{0.01\columnwidth}
\begin{picture}(100,50)
\put(-34,-17){\includegraphics[width=1.75\columnwidth]{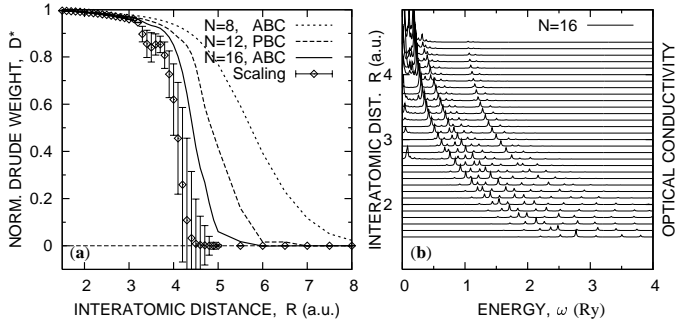}}
\end{picture}
\caption{Optical conductivity for the \emph{quarter--filled} chain: 
$(a)$ normalized Drude weight vs.\ lattice
parameter $R$ and its values extrapolated to $1/N\rightarrow 0$ 
(the Aitken method has been used to estimate the errors); 
$(b)$ regular part of the conductivity, $\si_{\rm reg}(\om)$ for 
$N=16$ atoms.}
\label{sigqf}
\end{figure}

The results for the 1D system of $N=6\div 14$ atoms in the \emph{half--filled} 
chain case are shown in Fig.\ \ref{sighalf}a. 
The values of $D_N^*$ used for the scaling (\ref{lndnscal}) are listed
in Table \ref{drudew:taba}, together with the resulting 
$D_{\infty}^*$ and its relative error (we stop at the lattice parameter $R$
for which $D_{\infty}^*=0$, within the range of errorbars).
We also provide, for comparative purposes, the data for the system with the
\emph{on--site} Coulomb interaction only (i.e.\ for the Hubbard model) in 
Table \ref{drudew:tabb}.
In both cases, i.e.\ that with the long--range interactions (\emph{cf.}\ Table 
\ref{drudew:taba}), and that with the on--site interaction only (\emph{cf.}\ 
Table \ref{drudew:tabb}), 
the extrapolated Drude weight $D_{\infty}^*$ becomes significantly greater than
zero (of $2\si$ value) only for small lattice parameter, i.e.\ for 
$R\leqslant 2.6a_0$ and $R\leqslant 2.1a_0$, respectively. 
The limiting values match well those for which the Luttinger--liquid 
exponent crosses the critical value $\theta=1$,  corresponding to the 
localization--delocalization boundary. 
The above results suggest again the localization onset in these 1D systems at 
half--filling. 
However, the optical conductivity $\si_{\rm reg}(\om)$, drawn in Fig.\ 
\ref{sighalf}b, shows the isolated Hubbard peak at $\om\approx U$ and 
\emph{no intraband transitions} present in the delocalized state. 
Because of this fact, and also because of the nonzero value of the charge--gap
for any $R$ (cf.\ \emph{above}), one should regard both the half--filled
systems studied here as the Mott insulators in the large $N$ limit.
Nevertheless, the finite--$N$ results show that the conductivity of the 
nanoscopic chain diminishes by two orders of magnitude when the corresponding 
increase of the lattice parameter $R$ is in the range of $40\div 50\%$. 
So, one can consider such system as undergoing a transformation either
from a \emph{nanoliquid} to the \emph{localized spin system} at the 
half--filling or from an intrinsically \emph{partially localized} state,
since the intraband transitions are absent even in the small--$R$ range.
It would be very interesting to confirm these results experimentally.

\begin{table*}[!t]
\caption{Normalized Drude weight $D_N^*$, the extrapolated value 
$D_{\infty}^*$, and its relative error for 1D half--filled system with 
long--range Coulomb interaction. The critical value $R_\crit\approx 2.6a_0$
is determined from the condition $D_{\infty}^*=2\si(D_{\infty}^*)$.}
\label{drudew:taba}
\begin{center}
\begin{tabular}{r|lllll|ll}
\hline\hline
$R/a_0$ & $D^*_{14}$ & $D^*_{12}$ & $D^*_{10}$ & $D^*_{8}$ & $D^*_6$
& $D^*_{\infty}$ & {\small $\si(D_{\infty}^*)/D_{\infty}^*$} \\ \hline
1.5 & 0.9225 & 0.9420 & 0.9563 & 0.9727 & 0.9822
 & 0.8008 & 0.019 \\
2.0 & 0.6245 & 0.7105 & 0.7875 & 0.8660 & 0.9222
 & 0.2567 & 0.129 \\
2.5 & 0.1839 & 0.2812 & 0.4109 & 0.5813 & 0.7526
 & 0.0087 & 0.420 \\
2.6 & 0.1269 & 0.2096 & 0.3315 & 0.5065 & 0.7009
 & 0.0033 & 0.489 \\
2.7 & 0.0840 & 0.1508 & 0.2595 & 0.4312 & 0.6441
 & 0.0012 & 0.566 \\
2.8 & 0.0536 & 0.1049 & 0.1972 & 0.3586 & 0.5836
 & 0.0004 & 0.641 \\
2.9 & 0.0315 & 0.0706 & 0.1456 & 0.2914 & 0.5208
 & 0.0001 & 0.920 \\ 
\hline
3.0 & 0.0196 & 0.0461 & 0.1047 & 0.2314 & 0.4575
 & 0.0000 & $\ \ \ -$ \\ 
\hline\hline
\end{tabular}
\end{center}
\end{table*}

\begin{table*}[!t]
\caption{Normalized Drude weight $D_N^*$, the extrapolated value 
$D_{\infty}^*$, and its relative error for 1D half--filled system 
described by the Hubbard model. The critical value $R_\crit\approx 2.1a_0$
is determined as for the long--range interaction case (\emph{cf.}\ Table 
\ref{drudew:taba}).}
\label{drudew:tabb}
\begin{center}
\begin{tabular}{r|lllll|ll}
\hline\hline
$R/a_0$ & $D^*_{14}$ & $D^*_{12}$ & $D^*_{10}$ & $D^*_{8}$ & $D^*_6$
 & $D^*_{\infty}$ & {\small $\si(D_{\infty}^*)/D_{\infty}^*$} \\ \hline
1.5 & 0.6173 & 0.6742 & 0.7342 & 0.7973 & 0.8640
 & 0.3294 & 0.070 \\ 
2.0 & 0.1529 & 0.2259 & 0.3276 & 0.4631 & 0.6331
 & 0.0092 & 0.364 \\ 
2.1 & 0.0991 & 0.1598 & 0.2527 & 0.3884 & 0.5731
 & 0.0032 & 0.449 \\ 
2.2 & 0.0619 & 0.1094 & 0.1896 & 0.3190 & 0.5123
 & 0.0010 & 0.540 \\ 
2.3 & 0.0372 & 0.0724 & 0.1382 & 0.2561 & 0.4514
 & 0.0003 & 0.639 \\ 
2.4 & 0.0218 & 0.0467 & 0.0985 & 0.2018 & 0.3927
 & 0.0001 & 0.739 \\ 
\hline
2.5 & 0.0124 & 0.0294 & 0.0687 & 0.1561 & 0.3372
 & 0.0000 & $\ \ \ -$ \\ 
\hline\hline 
\end{tabular}
\end{center}
\end{table*}

The situation again becomes completely different at the 
\emph{quarter--filling} (QF).
Namely, the normalized Drude weight of the QF systems of $N=8\div 16$ atoms, 
depicted in Fig.\ \ref{sigqf}a, shows a \emph{highly--conducting} behavior 
($D^*\!\approx\! 1$) for $R\lesssim 3.5a_0$, and gradually transforms to zero 
in the range $R/a_0=4\div 5$. 
Also, the regular part of the conductivity $\si_{\rm reg}(\om)$ (\emph{cf.}\ 
Fig.\ \ref{sigqf}b) comprises intraband transitions in the quasi--metallic 
range, particularly near the onset of the charge--density wave state 
(for $N=16$). 
Such a behavior provides the model case for the transformation from 
a nanometal in the small--$R$ range to the charge--ordered system 
(\emph{cf.}\ Fig.\ \ref{corfqf}) for larger $R$.

\section{A brief overview: Novel features}

We have provided a fairly complete description of the electronic states in 
a finite 1D chain within the framework of the EDABI method, which combines the 
\emph{exact diagonalization} of the many--fermion Hamiltonian in the Fock space
with a subsequent \emph{ab initio} readjustment of the single--particle 
(Wannier) functions.
Both the ground--state and dynamical properties have been obtained as a 
function of the variable lattice parameter (interatomic spacing) $R$.
Our approach thus \emph{extends} the current theoretical treatments of 
band and strongly correlated systems within the parametrized 
(second--quan\-ti\-zed) models, by determining those parameters and, in turn, 
analyzing the correlated state explicitly as a function of $R$.
The single--particle wave function is allowed to readjust in the correlated 
state within the EDABI method, thus unifiying the second and the first 
quantization aspects of the many--particle states into a single, 
\emph{fully microscopic} scheme.

We start by analyzing the situation with one valence electron per atom 
(the \emph{half--filled} case), and we include the \emph{long--range} Coulomb 
interaction. 
The Luttinger--liquid type of electron momentum distribution suggests
a \emph{crossover transition} from the quasi--metallic to the insulating 
(\emph{spin--ordered}) state with increasing $R$ (the same is true about 
the system without the long--range interaction \cite{thesis}, but the 
quasi--metallic behavior is  manifested to a much stronger degree when the 
long--range part of the Coulomb interactions is included). 
The finite--size scaling with ($1/N\rightarrow 0$), performed on the
charge--energy gap shows the insulating nature of the ground state for the 
large $N$ limit, in agreement with the renormalization--group results for the 
\emph{infinite} system with two Fermi points \cite{fabri}.
Such an apparently dychotomic nature (localized vs.\ itinerant) of the 
nanoscopic systems is confirmed by their transport properties.
On the one hand, the Drude weight is nonzero in the small $R$ limit, and the 
localization threshold agrees with those obtained from the Luttinger--liquid 
exponent, but the regular part of the optical conductivity exhibits the 
insulating behavior. This is the reason we coined the term: \emph{a partially 
localized quantum nanoliquid}. The most fundamental features of the electrons 
as a \emph{quantum nanoliquid} is provided in Figs.\ \ref{nksll}a and 
\ref{ak3d}b--d, where the \emph{Fermi--like} distribution, as well as the 
\emph{renormalized band structure} appear for small interatomic spacing. 
These features evolve with $R\rightarrow R_c$ into \emph{atomic--like} through 
the regime with split \emph{Hubbard--like} subbands.

An illustrative example of the nanonscopic system with a clear transformation
from \emph{nanometal} to \emph{nanoinsulator} with the charge--density wave 
order is provided with the \emph{quater--filled} nanochain (including again 
the long--range Coulomb interaction). 
For that system, the Drude weight is reduced gradually from its maximal value
to zero, and other properties evolve analogously with increasing lattice 
parameter $R$.
The intermediate range of $R$, where the crossover takes place, also shrinks
rapidly with increasing $N$, suggesting the existence of a sharp 
zero--temperature transition in the large $N$ limit. 

The above analysis, for both the half-- and the quarter--filled cases is 
very sensitive to the choice of boundary conditions. 
This problem is widely studied in the existing literature 
and its relation to the spontaneous magnetic flux 
appearing in the mesoscopic rings of $N=4n$ atoms
\cite{butic} has been established before \cite{abra}. 

The EDABI method implemented here can also be applied to discuss the coupling
of electrons to the lattice and the dimerization (for a discussion, see the 
third paper in Ref.\ \cite{spary}).
Also, one should incorporate the properties of the chain with an odd number of 
electrons (the boundary conditions will then involve the complex number 
domain). Furthermore, the dynamics with single holes and impurities in the 
nanochain would make the system more realistic as a quantum wire.
Finally, the application of the present scheme to higher ($ns$--like) 
valence orbital systems would make possible a direct comparison with 
experimental results on quantum nanowires made of noble and alkaline elements.
Nonetheless, we believe that our present analysis models the fundamental
features of such \emph{quantum nanowires}, particularly their 
quantum--nanoliquid
aspects (evolving Fermi--like distribution, renormalized bands with a gap 
even in the absence of a long--range order).
We should be able to see a progress along these lines soon.

\section*{Acknowledgement}
We thank our colleagues: Dr.\ Maciej Ma\'{s}ka and Prof.\ Krzysztof 
Ro\'{s}ciszewski, for discussions about the Lanczos algorithm and the role 
of boundary conditions for finite systems.
We are also grateful to the Institute of Physics of the
Jagiellonian University  for the support for computing facilities
used in part of the numerical analysis.
The support from the Committee for Scientific Research (KBN) of Poland
(Grant No.\ 2P03B~050~23), and from the Polish Foundation for Science (FNP), 
is acknowledged.
The authors are also gratefull to one of the Referee whose suggestions
improved substantially the paper presentation.
With this paper we would like to mark the European Union enlargement
on May 1, 2004.

%%%%%%%%%%%%%%%%%%%%%%%%%%%%%%%% BIBLIOGRAPHY %%%%%%%%%%%%%%%%%%%%%%%%%%%%%%%%%

%%%%%%%%%%%%%%%%%%%%%%%%%%%%%%%%%%%%%%%%%%%%%%%%%%%%%%%%%%%%%%%%%%%%%%%%%%%%%%
\end{document}